\documentclass[aps,showpacs,twocolumn,superscriptaddress]{revtex4-1}
\usepackage[utf8]{inputenc}
\usepackage{amsmath}
\usepackage{amssymb}
\usepackage{amsfonts}
\usepackage{makeidx}
\usepackage{graphicx}
\include{epsf}
\usepackage{bm}
\usepackage{clrscode}
\usepackage{color}

\newcommand{\lazo}[1]{\left\{ {#1} \right\} }
\newcommand{\mean}[1]{\left\langle {#1} \right\rangle }
\newcommand{\V}[1]{\bm{#1}}

\begin{document}

\title{Learning a Local Symmetry with Neural-Networks}
\author{A. Decelle}\affiliation{Laboratoire de Recherche en Informatique, TAU - INRIA, CNRS,  Universit\'e Paris-Sud et Universit\'e Paris-Saclay, B\^at. 660, 91190 Gif-sur-Yvette, France}
\author{V.~Martin-Mayor}\affiliation{Departamento  de F\'\i{}sica Te\'orica, Universidad Complutense, 28040 Madrid, Spain}\affiliation{Instituto de Biocomputaci\'on y F\'{\i}sica de Sistemas Complejos (BIFI), 50018 Zaragoza, Spain}
\author{B.~Seoane}\affiliation{Sorbonne Université, CNRS, IBPS, UMR 7238, Laboratoire de Biologie Computationnelle et Quantitative (LCQB), 75005
Paris, France}\affiliation{Sorbonne Université,  Institut des Sciences, du Calcul et des Données (ISCD), 75005 Paris, France}

\date{\today}

\begin{abstract}
We explore the capacity of neural networks to detect a symmetry with complex local and non-local patterns : the gauge symmetry $Z_2$. This symmetry is present in physical problems from topological transitions to QCD, and controls the computational hardness of instances of spin-glasses. Here, we show how to design a neural network, and a dataset, able to learn this symmetry and to find compressed latent representations of the gauge orbits. Our method pays special attention to system-wrapping loops, the so-called Polyakov loops, known to be particularly relevant for computational complexity. 
\end{abstract}


\maketitle

The physics community is now greatly excited by the possibilities offered by
machine learning tools, which have reached \emph{superhuman} performance in
tasks of significant complexity (think, for instance, of Go
playing~\cite{silver16gogame}). Indeed, deep (convolutional) neural networks
(DCNN)~\cite{lecun15deeplearning,schmidhuber15deeplearning}, initially
developed for classification and pattern recognition tasks, have been applied
to the identification of phases of
matter~\cite{torlai16machine,carrasquilla2017machine,wetzel2017machine,Nieuwenburg17confusion,wang16machine,ohtsuki17machine,beach18machineKT},
including glasses~\cite{schoenholz16machineglass,schoenholz17machineglass,cubuk15machineglass}
and topological states~\cite{deng17machinetopological}, or even to seemingly
for-humans-only tasks, such as finding real-space renormalization group
transformations~\cite{koch-janusz18machineRG} (this is just a somewhat
arbitrary selection of, literally, hundreds of applications to physics).

In this context, local -or gauge- symmetries pose a major challenge due to the
absence of any local or global order parameter~\cite{elitzur75}, which explains why only
preliminary studies have been
conducted~\cite{carrasquilla2017machine,wetzel2017machine} (we add to this
list). In fact, thanks to their convolutional layers, DCNN successfully handle locally
symmetries such as global translations and rotations: even if moved, DCNN still
identify a previously learned imaged. Therefore, the obvious next
step for Physicists is to consider more general symmetries for practical
purposes.

The specific question we had in mind was whether or not DCNNs could be used to
predict the {\em computational complexity} of a particular instance of an
optimization problem. Spin glasses represent the perfect playground to test
this idea, because finding the ground state of a simple Hamiltonian such
as: \begin{equation}\label{eq:EAmodel} \mathcal{H} =
  -\sum_{\mean{\V{x},\V{y}}} J_{\V{x}\V{y}} \sigma_{\V{x}}
  \sigma_{\V{y}}\,, (\sigma_{\V{x}}=\pm 1\ \text{for all sites}\ \V{x})\,,
\end{equation}
is an NP-complete problem as soon as the underlying interaction-graph is
non-planar~\cite{barahona1982on,istrail2000statistical} (we shall consider
statistically independent couplings $J_{\V{x}\V{y}}=\pm 1$ with $50\%$
probability). The classification problem is motivated because the
computational difficulty of solving different problem instances of
Eq.~\eqref{eq:EAmodel} spreads over several orders of
magnitude~\cite{janus:10,fernandez2013temperature,billoire:14,martin-mayor:15,fernandez2016temperature,billoire:18},
even for such a modest number of spins as $N\sim 500$~\footnote{Actually,
  Refs. \cite{janus:10,fernandez2013temperature,billoire:14,fernandez2016temperature,billoire:18}
  attempted to find equilibrium configurations using a Parallel Tempering
  algorithm down to some minimal temperature $T_{\mathrm{min}}$. In order to
  compute the Ground State, one needs to push $T_{\mathrm{min}}$ to zero, as
  done for instance in Ref. \cite{martin-mayor:15}. Unfortunately, the lower
  $T_{\mathrm{min}}$ the larger the spread over the samples of the
  computational hardness, see
  e.g. Refs. \cite{janus:10,fernandez2016temperature,billoire:18}.}.  In spite
of the question's practical relevance, it is still unknown which features of
the coupling matrix $J_{\V{x}\V{y}}$ cause this tremendous disparity of
computational cost~\cite{billoire:14}. DCNNs would be an obvious choice to
address the computational-cost classification problem, were it not for the
gauge symmetry of Hamiltonian~\eqref{eq:EAmodel} (the $\epsilon_{\V{x}}=\pm 1$
are arbitrary)~\cite{toulouse:77}
\begin{eqnarray}
 J_{\V{x}\V{y}}\to\tilde J_{\V{x}\V{y}} = J_{\V{x}\V{y}} \epsilon_{\V{x}} \epsilon_{\V{y}},&\text{ and }& \sigma_{\V{x}}\to
  \tilde\sigma_{\V{x}} = \epsilon_{\V{x}} \sigma_{\V{x}}\,.\label{eq:gauge}
\end{eqnarray}
All problem instances related by this transformation belong to the same
\emph{gauge orbit}. Now, the difficulty for solving problems from the same
gauge orbit is \emph{identical}. Hence, our dreamed DCNN should first be
able of telling us with certainty whether or not two problem instances belong
to the same gauge orbit. 

\begin{figure}
  \includegraphics[scale=0.39,trim=160 280 150 250,clip]{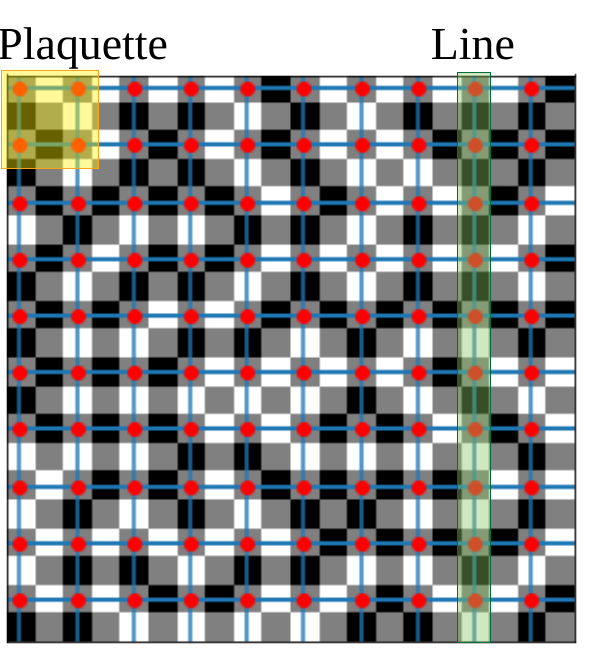}
  \includegraphics[scale=0.22,trim=20 110 20 150,clip]{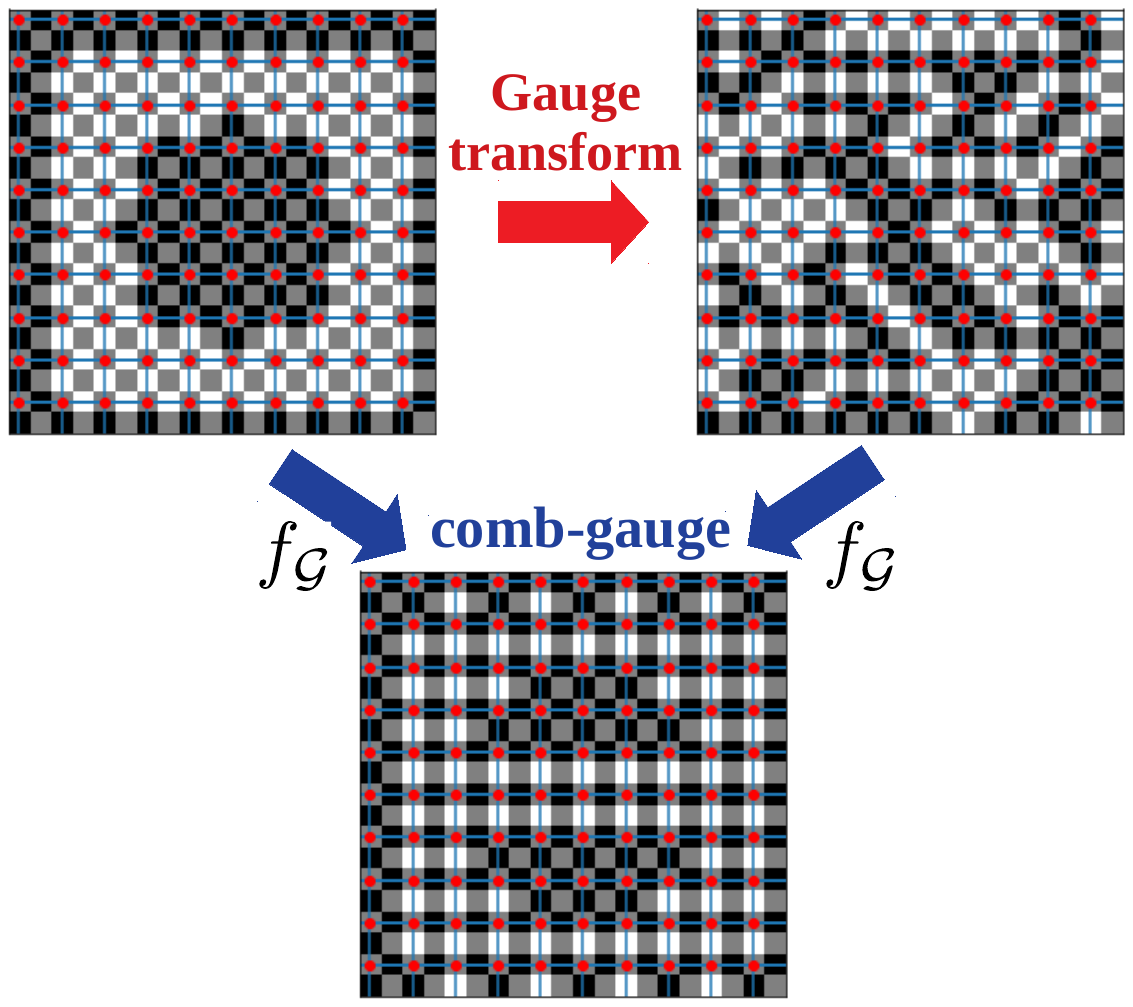}
  \caption{{\bf(Left)} Chess transformation from the lattice $(x,y)$,
    $0\leq x,y\leq L-1$, to the image ($x_1,x_2)$, $0\leq x_1,x_2\leq
    2L-1$. Periodic boundary conditions (PBC) are implemented by two
    additional rows and columns framing the image (for clarity, we
    only show the additional row at $x_2=-1$ and the additional column at
    $x_1=2L$). The spin at site $\V{x}=(x,y)$ is assigned to the pixel
    $(x_1=2x,x_2=2y)$ in the image, depicted as dummy gray cells,
    which are set to zero when fed to the neural network. The coupling
    $J_{\V{x},\V{y}}$ with $\V{x}=(x,y)$ and $\V{y}=(x+1,y)$ is in the
    pixel $(x_1=2x+1,x_2=2y)$ which is set to black if $J=1$ (white if
    $J=-1$). Similarly, the pixel $(x_1=2x,x_2=2y+1)$ contains the
    coupling between $(x,y)$ and $(x,y+1)$. The remaining pixels at
    the center of each plaquette, i.e. $(x_1=2x+1,x_2=2y+1)$, are also
    fixed as dummy gray pixels.  We indicate with red dots the
    spin-pixels per site, while the blue edges are in the $J$-pixels
    joining neighboring spin-pixels. We also show a plaquette and a
    Polyakov loop [the (say) vertical line, which is a closed loop
      thanks to the PBC].  {\bf(Right)} A problem instance and one of
    its gauge transforms. Both instances lead to the same comb-gauge
    representation after gauge-fixing.  } \label{fig_chess}
\end{figure}

Here we present a machine-learning algorithm that solves the problem
of gauge-orbit identification as formulated for spin glasses on the
square lattice. The same algorithm works in the cubic lattice,
although we are limited to systems of smaller linear size due to the
memory and computational costs.  Interestingly, all the standard DCNNs
for image classification tried, including the
ResNet~\cite{he2016deep}, completely failed at this task. A careless
posing of the problem could make it wrongly seem trivial. Indeed,
instances from the same orbit share the value of every Wilson
loop~\cite{montvay:97} [the product of couplings along a closed loop
  in the lattice, which is
  gauge-invariant~\eqref{eq:gauge}]. Attention immediately falls on
the \emph{plaquette}, the shortest Wilson loop, see
e.g.~\cite{carrasquilla2017machine} or
Fig.~\ref{fig_chess}--left. However, two instances sharing the value
of \emph{every} plaquette, but differing on the so-called Polyakov
loops (the shortest Wilson loops wrapping the system thanks to the
periodic boundary conditions), may have vastly different computational
complexity~\cite{fernandez2016temperature}.  We improve over
Ref.~\cite{carrasquilla2017machine} by teaching our machine to
consider both local and non-local Wilson loops when studying a $Z_2$
gauge symmetry.

Let us highlight two other aspects of this problem that machine-learning
practitioners may find attractive: (i) a training set of (essentially) arbitrary
size can be easily generated and (ii) an algorithm of polynomial complexity
provides an exact answer to the question of whether two problem
instances belong to the same gauge orbit.

Below, we present two different approaches to solve this classification
problem using DCNN (we employed the Keras-tensorflow and scikit-learn
libraries~\cite{chollet2015keras, scikit-learn}). Our first algorithm tells us
if two problem instances are in the same gauge orbit. Our second algorithm is
an autoencoder, a DCNN capable of finding a latent representation of a gauge
orbit by means of an approximate gauge-fixing. Although the latent
representation can be used for classification purposes as well, its strength
is in that it clusters problem instances by orbits.

For square lattices, it is natural to feed the coupling matrix $\V{J}$
to the neural network as an image. After considering several
alternatives, our choice was to map our physical square lattice of
size $L$ to a square image of size $2L$ through the \emph{chess}
transformation illustrated in Fig. \ref{fig_chess}--left (the
chess-transformation generalizes to 3D). Although one pixel out of two
is wasted in the resulting image, we found that the learning process
and the interpretation of results were easier with the chess
transformation than with less memory-demanding representations.

 Gauge transformations are also illustrated in Fig. \ref{fig_chess}--right:
 the naked eye can hardly tell whether or not the images corresponding to two
 coupling-matrices belong to the same gauge orbit. This question can be
 answered by fixing the gauge~\footnote{In this work we deal with an Abelian
   gauge group which makes fixing the gauge simple (difficulties arise for
   non-Abelian gauge groups, see e.g. Ref.~\cite{marinari:91}).}, that is, to
 use a map $f_\mathcal{G}:\boldsymbol{J}^{\mathcal{O}_k}\to { \boldsymbol{\hat
     J}}^{\mathcal{O}_k}$ from any instance $\boldsymbol{J}$ from gauge orbit
 ${\mathcal{O}_k}$ to a single representative of it, $\boldsymbol{\hat
   J}$. Thus, two instances are in the same orbit if, and only if,
 $f_\mathcal{G}(\boldsymbol{J})=f_\mathcal{G}(\boldsymbol{J^\prime})$.  We
 construct our mapping by changing the gauge: the
 $\boldsymbol{\epsilon}\equiv\lazo{\epsilon_{\V{x}}}$ in Eq.~\eqref{eq:gauge}
 are chosen in such a way that $\tilde J_{\V{x},\V{y}}=1$ for any horizontal
 coupling $\V{x}-\V{y}=(\pm 1,0)$ (but for $\tilde
 J_{\V{x}=(L-1,y),\V{y}=(0,y)}$ which is equal to a gauge-invariant
 Polyakov loop), as well as $\tilde
 J_{\V{x}=(0,y),\V{y}=(0,y+1)}=1$ for $0\leq y<L-2$.  We include a code
 performing this gauge-fixing in the Appendices.


{\bf Construction of the data set--} We found inconvenient for our
purposes the approach used in Ref.~\cite{carrasquilla2017machine} to
detect the gauge symmetry, namely constructing a (balanced) dataset of
pairs of systems, a group with pairs of instances from the same orbit
an the other group with pairs of randomly-chosen
$\boldsymbol{J}$s. Indeed, this classification problem is too
easy. Most of the time, and this is what the DCNN will learn, the pair
of randomly-chosen $\boldsymbol{J}$s will be so different that one
could tell that they do not belong to the same orbit just by looking
at a very reduced number of plaquettes~\footnote{For two
  randomly-chosen $\boldsymbol{J}$s, the probability of coincidence in
  $k$ fixed, non-overlapping plaquettes falls as $1/2^k$.}. A DCNN
trained in this way would completely miss situations in which just a
few coupling changed, and it would be blind to extensive
transformations that leave every plaquettes unaltered.  Therefore, we
need to ensure that in our dataset it will not be enough for the DCNN
to check one (or few) plaquette(s) [neither fixed plaquettes nor
  randomly chosen ones].

Specifically, our data-set is composed of $N_\mathrm{s}$ pairs
$\{\boldsymbol{J},\boldsymbol{J}'\}$. The $\boldsymbol{J}$ is random (with
uniform distribution). For half of the $N_\mathrm{s}$ pairs,
$\boldsymbol{J}'=\boldsymbol{J}$. In the other half, $\boldsymbol{J}'$ is
obtained from $\boldsymbol{J}$ by some transformation (see below and Appendices) that
changes only a small fraction of the couplings $J_{\V{x},\V{y}}$.  For all
pairs, $\boldsymbol{J}'$ is gauge-transformed (with random
$\lazo{\epsilon_{\V{x}}}$) before being fed to the DCNN.

In the so-called $\boldsymbol{J}'= R_q(\boldsymbol{J})$
transformation, a fraction $q$ of randomly-chosen $J_{\V{x}\V{y}}$ is
flipped. 

In the (horizontal) line-transformation $\boldsymbol{J}'=
L(\boldsymbol{J})$, $\boldsymbol{J}'$ is obtained from
$\boldsymbol{J}$ by flipping the couplings joining $\V{x}=(0,y)$ and
$\V{y}=(1,y)$ for any $y$ (vertical transformation: $\V{x}=(x,0)$ and
$\V{y}=(x,1)$, for all $x$). Every plaquette in the lattice take the
same value in $\boldsymbol{J}$ and $\boldsymbol{J}'$, but the sign of
all their horizontal (vertical) Polyakov loops is opposite.  These
line transformations~\footnote{Any other transformation can be
  expressed as a combination of broken plaquette(s) and/or line(s).},
are important when assessing the computational
hardness~\cite{fernandez2016temperature}.

In our data set, we choose with 1/3 probability
$\boldsymbol{J}'=L(\boldsymbol{J})$ or, with probability $2/3$,
$\boldsymbol{J}'=R_{q}(\boldsymbol{J})$. Line transformations are
equally likely to be horizontal or vertical. If the chosen
transformation is $R_q$, in order to force the scan of every
plaquette, we pick $q\sim 1/L^2$ with $50\%$ probability (we invert randomly
$1\!-\!5$ couplings), or $q=q_R$ where $q_R$ is an
uniform random number with $1/(2L^2)\le q_R<1/4$.

\begin{figure}
  \centering
  \includegraphics[scale=0.39,trim=80 10 70 90 ,clip]{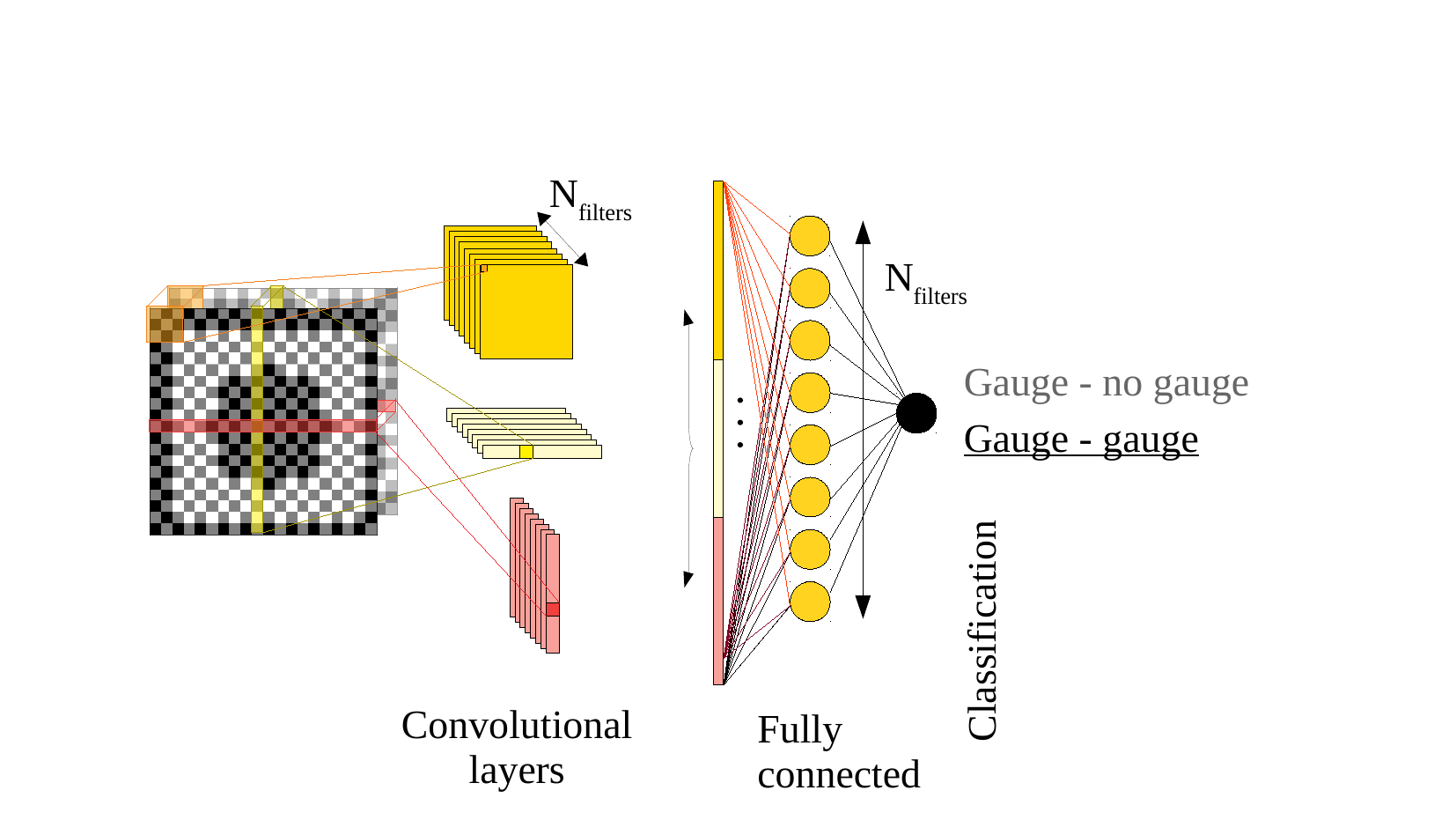}
  \includegraphics[scale=0.55,trim=30 550 30 100 ,clip]{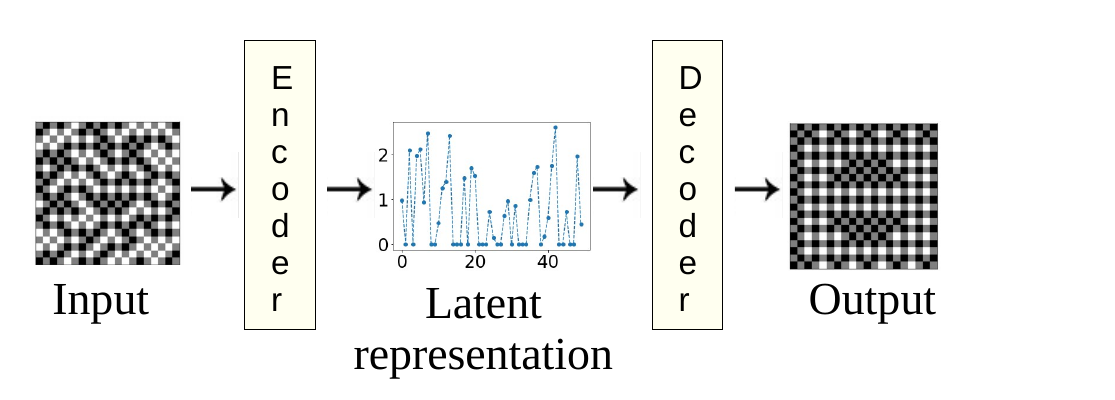}
  \caption{({\bf Top}) The typical architecture used for detecting the
    gauge symmetry between pairs of systems. It is important to scan
    both square-like kernels for the plaquettes and full-line kernels
    for the Polyakov loops. ({\bf Bottom}) Schematic representation of
    the autoencoder. The encoder is very similar to the architecture
    above, the decoder is typically made of upsampling layers
    (increasing the size of the input) and of convolutional
    layers.} \label{fig_DCNN}
\end{figure}
{\bf Construction of the DCNN--} We aim to build a DCNN that inputs
the chess-transformed (see Fig. \ref{fig_chess}--left) images
representing a pair of coupling matrices
$\{\boldsymbol{J},\boldsymbol{J}'\}$ and outputs the probability that
the two instances belong to the same gauge-orbit.

The Euclidean geometry of our problem suggests to use convolutional
neural networks (CNN)
\cite{fukushima1980neocognitron,lecun1989backpropagation,krizhevsky2012imagenet},
which are well adapted to translational
symmetry. Specifically, we combine in parallel three CNNs that scan simultaneously the
plaquettes, (square in Fig.~\ref{fig_DCNN}--top), and the Polyakov
loops, scanned through horizontal and vertical $1\times L$ slabs
(rectangles in Fig.~\ref{fig_DCNN}--top). The first CNN allows us to
find quickly small defects in the gauge symmetry, while the other two
search for non-local defects. These three CNNs serve as feature
detectors before a fully-connected layer that performs the
classification. We illustrate on Fig. \ref{fig_DCNN}--top the general
architecture of our DCNN (the number of layers and the size of the
dense layer vary with $L$). Additional details, as well as sample
programs, can be found in the Appendices.

{\bf Results for the classifying DCNN--} For our data set, we manage to
obtain almost $100\%$ of accuracy on linear sizes of $L=5,10$. In
other words, even for our very exigent data set, the DCNN learns to
tell whether or not two problem instances really are the same problem
in disguise.

\begin{figure}
  \includegraphics[scale=0.45]{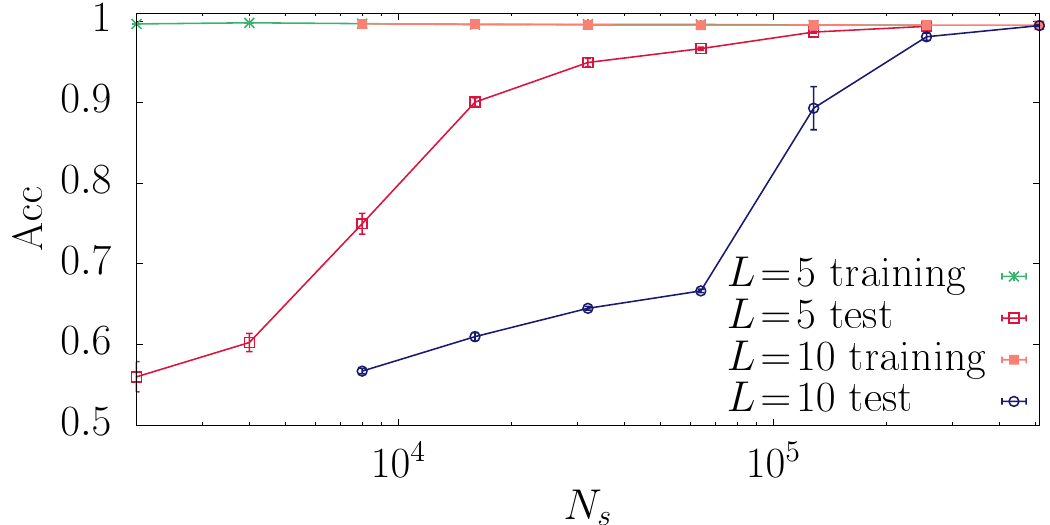}
  \caption{General accuracy of the classification task of the pairs of
    samples in our dataset (both for the training and the test set),
    as computed for lattices of sizes $L=5$ and $10$. Data and errors
    are computed from averages over 5 independent learning runs and
    datasets.} \label{fig_acc}
\end{figure}

However, let $N_s(p)$ be the size of the training set needed to reach
a target accuracy $p$. We see in Fig.~\ref{fig_acc} that $N_s(p)$ is
much smaller in the training set that in the test set (problem
instances in the test set are new to the DCNN).  Furthermore, $N_s(p)$
grows significantly with $L$.

We have found that the difficulty of the problem is largely caused by
the Polyakov-loop flipping line-transformations. More details on this
analysis can be found in the Appendices.

{\bf Learning to fix the gauge--} Gauge-fixing may be regarded as an
algorithm to reduce the dimensionality of the coupling matrix $\V{J}$
with no information loss. Hence, it is natural to ask ourselves if a
particular type of DCNN, an auto-encoder
(AE)~\cite{rumelhart1985learning,ballard1987modular}, may learn to fix
the gauge. Indeed, an AE takes an input vector $\V{x}$ and maps it to
a latent representation $f_\mathcal{E}(\V{x})$ (typically,
$f_\mathcal{E}(\V{x})$ is of smaller dimensionality than $\V{x}$). A
decoder generates a reconstructed vector from the latent
representation afterwards, $\V{x}' =
f_\mathcal{D}(f_\mathcal{E}(\V{x}))$. The weights of the encoder
$f_\mathcal{E}$ and the decoder $f_\mathcal{D}$ functions are chosen to
minimize a loss function (e.g. the $L_2$ distance between
$\V{x}$ and $\V{x}'$).

At variance with the traditional approach, we will not ask our AE to
reconstruct the input but to fix the gauge, that is to reconstruct a
unique $\boldsymbol{\hat J}$ (the comb gauge described above) for all
the instances in a given gauge orbit.

Our encoder will essentially share the architecture of our classifying
DCNN (namely, the three CNNs of Fig.~\ref{fig_DCNN} without the
classification layer). The decoder takes the encoder's output, and
pipes it to an upsampling layer, followed by our three feature
detector CNNs and by a last CNN from which we take the output (more
details can be found in the Appendices).  The output from a given coupling
matrix $\boldsymbol{J}$ is an attempted reconstruction of its
comb-gauge representation (Fig.~\ref{fig_chess}--right).

The AE can be used as a classifier simply by comparing the ``comb-gauge''
obtained from two problem instances. As shown in Table \ref{tab:gt},
only pairs of instances from the same orbit have a similar ``comb-gauge''
(the performance does not deteriorate when the system size increases).
\begin{table}[b]
\begin{tabular}{c|c|c|c|c|c|c}
$L$ &$N_\mathrm{s}$  &$N_\mathcal{O}$ &$p^{J,J}$  & $p^{J,J^\prime}$& $p^{J,R_{q=0.1}(J)}$& $p^{J,L(J)}$\\
\hline
$5$ & 100k & 1k &$\sim\!3 \%$ & $\sim\!52\%$ & $\sim\!30\%$ & $\sim\!21\%$ \\
$6$ & 400k & 1k &$\sim\!3.5\%$ & $\sim\!54\%$ & $\sim\!27\%$ & $\sim\!17.5\%$ \\
$8$ & 800k & 4k &$\sim\!3.1\%$ & $\sim\!52\%$ & $\sim\!30\%$ & $\sim\!10.5\%$
\end{tabular}
\caption{{\bf The autoencoder as a classifier.} Fraction of
  not-trivially-one couplings that are different in the ``comb-gauge''
  output of the AE as applied to two instances $\{\V{J},\V{J}'\}$
  from: $\V{J}=\V{J'}$ [$p^{J,J)}$], $\V{J}'=R_{q=0.5}(\V{J}')$
  [$p^{J,J^\prime}$], $\V{J}'=R_{q=0.1}(\V{J})$ [$p^{J,R_q(J)}$] or
  $\V{J}'=L(\V{J})$ [$p^{J,L(J)}$].  $\V{J}'$ is gauge-transformed
  (with random $\V{\epsilon}$) previously to the AE analysis. The AE
  was trained with $N_S$ instances, randomly extracted from $N_0$
  orbits. The results were computed from 1000 pairs
  $\{\V{J},\V{J}'\}$, with $\V{J}$ extracted from orbits not in the training
  set.} \label{tab:gt}
\end{table}

\begin{figure}
  \includegraphics[scale=0.28,trim=57 50 45 20,clip]{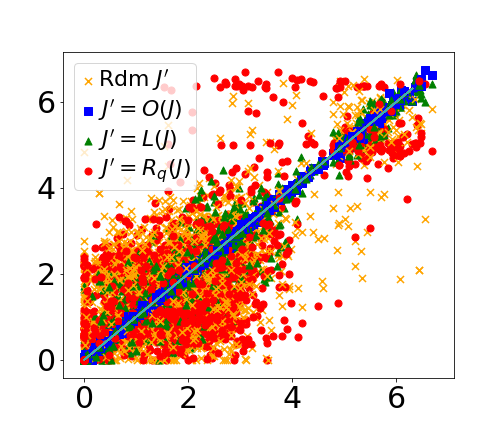}
  \includegraphics[scale=0.249, trim=78 50 62 55,clip]{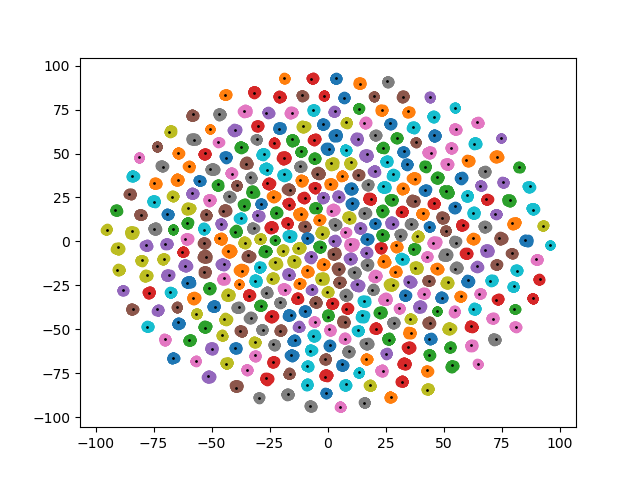}
  \caption{Visualizing the 50-dimensional autoencoder's latent
    representation. ({\bf Left}) scatter-plot comparison for pairs of
    problem instances $\{\V{J},\V{J}'\}$.  We display 50 points for
    each pair $\{\V{J},\V{J}'\}$, namely $(x_i^{\V{J}},x_i^{\V{J}'})$
    where $x_i^{\V{J}}$ and $x_i^{\V{J}'}$ are the $i$-th coordinates
    of both latent representations. We consider pairs
    $\{\V{J},\V{J}'\}$, $\V{J}'$ is randomly gauge-transformed
    previously to the AE analysis, with: $\V{J}'=\V{J}$ (blue squares),
    $\V{J}'=R_{q=0.5}(\V{J})$ (orange crosses), $\V{J}'=R_{q=0.1}(\V{J})$
    (red circles) and $\V{J}'=L(\V{J})$ (green triangles). The plot contains data from
    50 pairs of each type. ({\bf Right}) two-dimensional t-sne
    representation~\cite{maaten2008visualizing} of the latent
    representation as obtained for 20000 instances randomly extracted
    from 200 (unrelated) gauge orbits. Instances from the same orbit
    are represented with the same color (some orbits share color, due
    to our limited palette). Instances from the same orbit
    cluster. The black points at the center of each cluster are the
    t-sne coordinates for the latent representation obtained for the
    gauge-comb representative of each of the 200
    orbits.} \label{fig:tsne}
\end{figure}

We can gain some understanding by visualizing the latent
representation, see Fig.~\ref{fig:tsne}. Indeed the AE's latent
representation clusters problem instances belonging to the same
orbit. Furthermore, not only the representation for two problems from
the same orbit is nearly identical: changing a few links or performing
a line transformation results into a significantly different latent
representation.

{\bf Conclusions--} We have demonstrated a successful machine learning
approach to detect whether or not two spin-glass instances are
mutually related by a gauge transformation. This problem is
particularly challenging for neural networks due to the absence of an
order parameter. In fact, we have checked the failure of the standard DCNNs for image classification, such as pre-trained DCNNs, no matter the size of the training
set.  Our results underline the necessity of carefully choosing the
learning dataset, if we want the DCNN to learn the full symmetry
(which includes global Wilson loops). We show that our DCNNs are able
to learn the gauge symmetry and even to find a latent representation
that can be used to fix the gauge.  This success comes at the cost of
very large training datasets, whose size need to grow with the system
size. Now that we have in our hands DCNNs able to identify gauge
symmetries, we will approach our original question, namely \emph{what
  makes certain problem instances far more computationally costly than
  others?}

\begin{acknowledgments}
We thank L. A. Fern\'andez for encouraging discussions and Marco
Baity-Jesi for his careful reading of the manuscript.  This work was
partially supported by Ministerio de Econom\'ia, Industria y
Competitividad (MINECO) (Spain) and by EU's FEDER program through
Grant No. FIS2015-65078-C2 and by the LabEx CALSIMLAB (public grant
ANR-11-LABX-0037-01 constituting a part of the ``Investissements
d'Avenir" program - reference : ANR-11-IDEX-0004-02).
\end{acknowledgments}

\newpage

\onecolumngrid
\appendix

\section{Sample generation and basic transformations} \label{sec:transformation}
The first step to build our dataset is to create independent realizations of the disorder  $\boldsymbol{J}$ (what we call sample). The generation codes for all the functions mentioned below can be downloaded from file \texttt{src/tools.py} in Ref.~\cite{git-ref}.

\begin{itemize}
\item {\bf Generation of a random sample $\boldsymbol{J}$:} A random sample $\boldsymbol{J}$ is generated by assigning a random sign ($\pm 1$) to each of the  $2 L_x L_y$ couplings in the two dimensional lattice system. The code to create a sample can be found in function  \texttt{createSample\_2D}. 
\end{itemize}

\begin{figure}[h!]
  \includegraphics[scale=0.6,trim=0 40 0 0 ,clip]{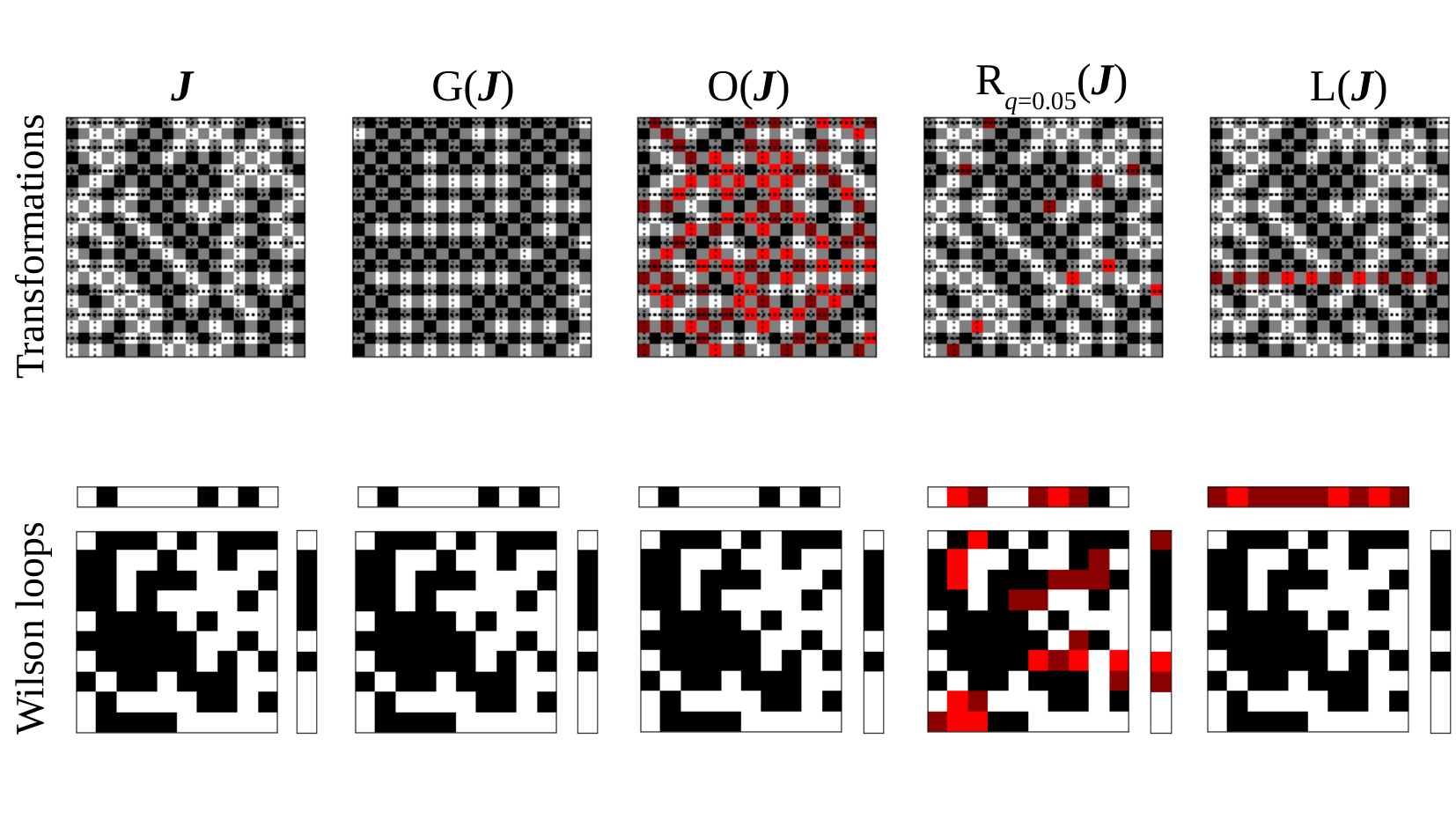}
  \caption{{\bf Top row}: a sample $\boldsymbol{J}$ and the possible
    transformations used to build our dataset. {\bf Second row:} for
    each of the samples depicted in the top-row, we show the sign of
    all the plaquettes (i.e. product of the $J_{\V{x},\V{y}}$ along
    the plaquette) in the system and the sign of all the Polyakov
    loops (i.e. product of the $J_{\V{x},\V{y}}$ along the horizontal
    and vertical lines; Polyakov loops are represented outside of the
    square).  We highlight in red the couplings (top row) or the
    Wilson lops (bottom row) that change, as compared with the
    leftmost image. For a given sample $\boldsymbol{J}$, a random
    gauge-transform changes approximately 50\% of the couplings but no
    plaquette or Polyakov loop.  On the other hand, changing just a few
    couplings (see $\mathrm{R}_{q=0.05}$ where $5\%$ of the couplings were
    flipped), has strong effects both in the plaquettes and the
    Polyakov loops. The last column shows that, flipping a full
    vertical line of couplings does not break any plaquette: this
    transformation can only be detected in the Polyakov
    loops.} \label{fig_fgauge}
\end{figure}

In addition, we  consider 4 possible transformations of these samples (all of them are illustrated in the first row of Fig.~\ref{fig_fgauge}):
\begin{itemize}
\item {\bf Gauge fixing $\mathrm{G}(\boldsymbol{J})$:} we map our
  sample $\boldsymbol{J}$ to its comb-gauge representative. To do so,
  we use the gauge transformation explained in Eq. (2) of the
  main-text. Specifically, we fix to one (black in our color code) all
  the couplings in the horizontal direction, as well as the couplings
  in the first vertical column. However, the last coupling along each
  direction cannot be fixed due to the boundary conditions. The code
  to fix the gauge can be found in function \texttt{
    gauge\_fixing\_Comb}.
\item {\bf Random orbit $\mathrm{O}(\boldsymbol{J})$:} We use Eq. (2)
  of the main-text to generate a random representative of the
  gauge-orbit to which $\boldsymbol{J}$ belongs. Specifically, we
  generate $L_x L_y$ random signs $\epsilon_{\boldsymbol{x}}$ and set
  $ J^\prime_{\V{x}\V{y}} = J_{\V{x}\V{y}} \epsilon_{\V{x}}
  \epsilon_{\V{y}}$.  The code that performs the random-orbit
  transformation can be found in function \texttt{getOrbit\_2D}.
\item {\bf Random flip-coupling - $\mathrm{R}_q(\boldsymbol{J})$:} We
  invert the sign of a fraction $q$ of the $2 L_x L_y$ couplings
  in the system. The corresponding code can be found in function \texttt{
    getRandom\_2D}.
\item {\bf Line transformation - $\mathrm{L}(\boldsymbol{J})$:} We invert the sign of an horizontal or vertical line of non-connected couplings (see Fig.~\ref{fig_fgauge}). The code to generate this transformation is in function \texttt{getLine\_2D}.
\end{itemize}
One can consider more general transformations, like flipping a random connected line (not necessary straight) or a random loop of couplings in the system (codes can be found in  \texttt{getRandomLine} and \texttt{getLoop} functions). All them can be decomposed as a combination of the previous 4 transformations. We did not find any particular advantage to include them in the dataset for the learning, but we  checked that our trained machine classifies them correctly.

In order to distinguish between transformations that conserve the
gauge orbit [here, $\mathrm{G}(\boldsymbol{J})$ and
  $\mathrm{O}(\boldsymbol{J})$], from those that modify the orbit
[namely, $\mathrm{R}_q(\boldsymbol{J})$ and
  $\mathrm{L}(\boldsymbol{J})$], one needs to compute the Wilson
loops, as shown in Fig.~\ref{fig_fgauge}).  In particular, we note
that the $\mathrm{L}(\boldsymbol{J})$ transformation is particularly
difficult to detect since this transformation conserves all the
plaquettes, and the broken loops can be only detected through the
Polyakov loops.

\section{Additional details on the classifier DCNN gauge/not gauge} \label{sec:class}

The classifier aims to classify whether or not pairs of samples
$\{\boldsymbol{J}$, $\boldsymbol{J^\prime}\}$ belong to the same gauge
orbit. We begin with the construction of our dataset. 
\subsection{Dataset}\label{sec:simple-dataset}

We consider $N_\text{s}=2 M$ pairs $\{\boldsymbol{J}$,
$\boldsymbol{J^\prime}\}$.   In all cases, the original sample $\{\boldsymbol{J}\}$
is chosen randomly (with uniform probability). We refer to
Section~\ref{sec:transformation} for the definition of the
transformations.
\begin{itemize}
    \item {\bf Class 1:} $M$ pairs are taken from the same gauge
      orbit, $\boldsymbol{J^\prime}=\mathrm{O}(\boldsymbol{J})$.
    \item {\bf Class 2:} $M$ pairs of samples $\{\boldsymbol{J}$,
      $\boldsymbol{J^\prime}\}$ belonging to two different gauge
      orbits. This dataset is constructed as follows:
    \begin{itemize}
        \item {\bf Quite different orbits {\bf G1}:} $M/3$ pairs with
          $\boldsymbol{J^\prime}=\mathrm{O}\left
          (\mathrm{R}_q(\boldsymbol{J})\right)$) with $q\in [1/(2L_x
          L_y),0.25]$. This class ranges covers from samples with just
          one link flipped, to samples $\boldsymbol{J^\prime}$ where
          (almost) every plaquette has a chance to flip ($q=0.25$).
        \item {\bf Extremely similar orbits {\bf G2}:} $M/3$ pairs
          with $\boldsymbol{J^\prime}=\mathrm{O}\left
          (\mathrm{R}_q(\boldsymbol{J})\right)$) and $q\in [1/(2L_x
            L_y),5/(2L_x L_y)]$, so that only 1 to 5 links were
          inverted. Our motivation for introducing this group was
          forcing the machine to check \emph{every} plaquette in the
          system.
        \item {\bf Broken lines {\bf G3}:} $M/3$ pairs with $\boldsymbol{J^\prime}=\mathrm{O}\left (\mathrm{L}(\boldsymbol{J})\right)$). The line is horizontal or vertical with 50\% of the probability.
    \end{itemize}
\end{itemize}
An example of the generation of this dataset can be found in the notebook ~\texttt{DCNN\_simple.ipynb} in Ref.~\cite{git-ref}.

\subsection{Network} The structure of the neural network is illustrated in  Fig.~\ref{fig_DCNN}. We include the technical details of the network used in Table~\ref{tab:archDCNN}. We use the same architecture for all the $L$ and $N_{\mathrm{s}}$ discussed in the main-text. We include an example of the program used in \texttt{DCNN\_simple.ipynb} in Ref.~\cite{git-ref}.


\begin{table}
\centering
\begin{tabular}{cccccc}
\hline
LayerName & Input & LayerType & Activation & Nb of units & Kernel \\
\hline
\textbf{\underline{Simple DCNN}} \vspace{0.2cm} \\ 
CSq1 & $\V{J}$ & conv & ReLu & $64$ & $ 3 \times 3$ \\
CLh & $\V{J}$ & conv & ReLu & $64$ & $ L_x \times 1$  \\
CLv & $\V{J}$ & conv & ReLu & $64$ & $ 1 \times L_y$ \\
Dense1 & [CSq1,CL1,CC1] & FF & ReLu & $64$ & \\
Dense2 & Dense1 & FF & sigmoid & $1$ & \\
\hline
\end{tabular}
\caption{Architecture used for the simple classifier for gauge-not gauge pairs; conv stands for convolutional and FF for feed-forward.}
\label{tab:archDCNN}
\end{table}


In order to avoid overfitting, and also to avoid getting stuck in not optimal minima during the learning process, we found useful to alternate between two optimizers, in particular, between stochastic gradient descent and Adam \cite{adam}. An example of the strategy followed can be found in Ref.~\cite{git-ref}.

\subsection{Tests on the different groups of the dataset}

\begin{figure}[!h]
\centering
  \includegraphics[width=\columnwidth]{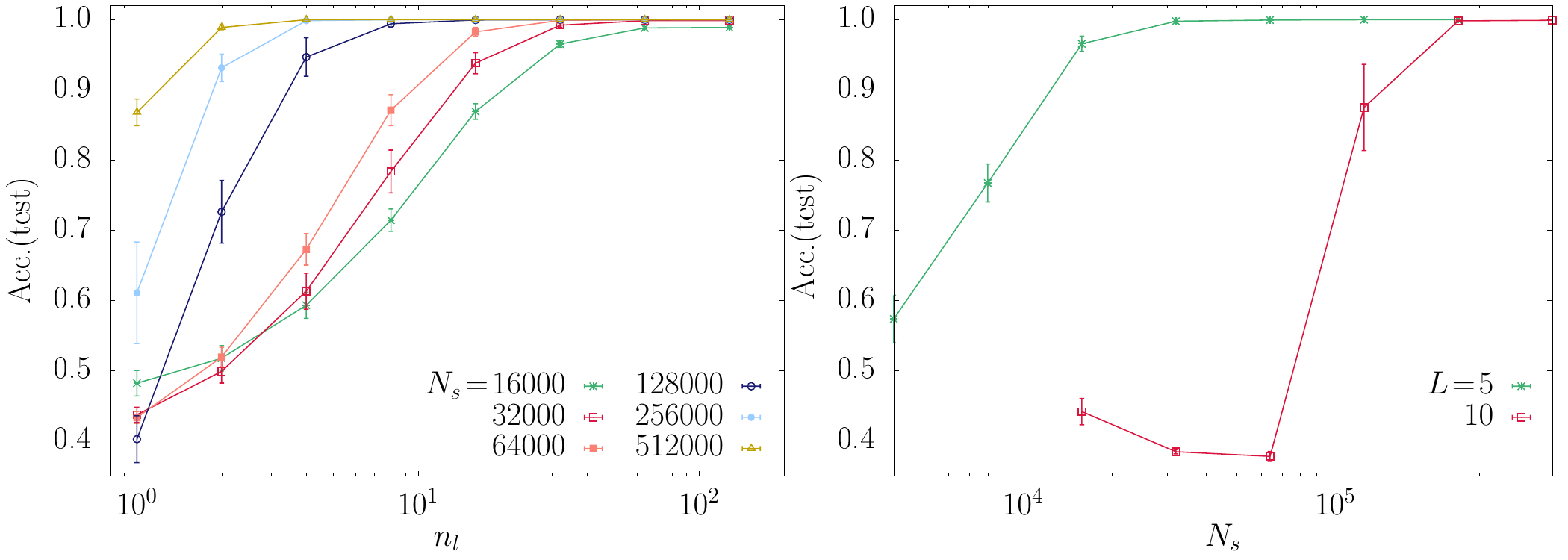}
  \caption{Accuracy performance, as extracted from the test dataset.
    The accuracy measures the probability that the machine correctly
    classifies as \emph{not from same-orbit\/} a pair
    $\{\boldsymbol{J}$, $\boldsymbol{J^\prime}\}$ with
    $\boldsymbol{J^\prime}=\mathrm{O}\left
    (\mathrm{R}_q(\boldsymbol{J})\right)$ (left panel) or
    $\boldsymbol{J^\prime}=\mathrm{O}\left
    (\mathrm{L}(\boldsymbol{J})\right)$ (right panel). {\bf(Left)}
    Accuracy of the classification for lattices $L=10$ (hence
    containing 200 couplings) as a function of the number of couplings
    $n_\mathrm{l}$ inverted by the $\mathrm{R}_q$ transformation. Data joined
    with lines were obtained with machines trained with the same
    number of pairs, $N_\mathrm{s}$. We see that the size of the
    training set needed to reach any accuracy threshold (0.95, say)
    rises dramatically upon decreasing $n_\mathrm{l}$. {\bf(Right)}
    The figure shows the accuracy, as computed from pairs in the test dataset
    with $\boldsymbol{J^\prime}=\mathrm{O}\left
    (\mathrm{L}(\boldsymbol{J})\right)$, versus the size of the
    training set $N_\mathrm{s}$. We show data for $L=5$ and $10$. Data
    and errors are computed from averages over 5 independent learning
    runs and datasets.} \label{fig_acc2}
\end{figure}

Fig.~\ref{fig_acc} shows the overall accuracy of DCNN
classifier, making no distiction about the {\bf G1}, {\bf G2} and {\bf
  G3} groups in Section~\ref{sec:simple-dataset}. We provide this
information, as obtained from pairs of samples in the test dataset, in
Fig.~\ref{fig_acc2}. In particular, a comparison of
Fig.~\ref{fig_acc2}--right (which corresponds to the line-transformed
samples in group {\bf G3}) with Fig. 3 in the main-text will convince
the reader that the global accuracy of the machine is dominated by
this group.

\section{Additional details on the autoencoder DCNN}

The autoencoder aims to find a latent representation of the
gauge-orbit by relating any sample to an unique representative of its
gauge-orbit (namely the comb-gauge representative). With this purpose
in mind, we built our dataset as explained in the next paragraph.

\subsection{Dataset} We will consider separately $N_\mathrm{g}$ distinct gauge orbits, identified by one orbit representative. We construct the orbits in the following way:
\begin{itemize}
\item $N_\mathrm{g}/2$ are generated as random samples $J$
(the probability that two random samples belong to the same orbit is negligible).
We call this set $\mathcal{R}_\mathrm{g}$.
\item $N_\mathrm{g}/4$ orbits were constructed by randomly selecting
  one $\V{J}$ from set $\mathcal{R}_\mathrm{g}$, and then setting as
  orbit-representative $\mathrm{R}_q(\boldsymbol{J})$, with $q$ an uniform random number $q \in [1/(2L_xL_y),0.25]$. 
\item $N_\mathrm{g}/4$ orbits were constructed by randomly selecting
  one $\V{J}$ from set $\mathcal{R}_\mathrm{g}$, and then setting as
  orbit-representative $\mathrm{L}(\boldsymbol{J})$.
\end{itemize}
We extract $N_\mathrm{s}$ distinct samples from each orbit by using
the $O$ transformation, recall Section~\ref{sec:transformation}.  An
example of the generation of this dataset can be found in the notebook
~\texttt{AutoEncoder.ipynb} in Ref.~\cite{git-ref}.

\subsection{Network} 

The \textbf{encoder} is typically built upon the model from the
main-text, see Fig. 2. The number of filters used for the
convolutional layers do not need to be very high.  For instance, $16$
filters are enough for a small lattice size (e.g. $L=5$). The results
of the three parallel CNNs are concatenated and then connected to a
dense network of size $L\times L \times N_{\rm latent}$, where $N_{\rm
  latent}$ is adjusted depending on the system size (we remind here
that the input of the encoder is of size $2L \times 2L$ because of the
chess transformation). The \textbf{decoder} is then made of, first, a
CNN and an upsampling layer in order to go back to the correct lattice
size. Then again, our three parallel CNNs are stacked (square,
vertical and horizontal kernel), taking as input the output of the
upsampling layer. Their outputs are concatenated before a last CNN
with a larger kernel (typically half of the system size). All the
parameters here can, of course, be adjusted to obtain the best result
possible for a given $L$. However, in front of the wide variety of
possible working parameters, we stuck to the above ones because
changing parameters did not result into a great improvement. In table
\ref{tab:archAE} we show an example of the architecture used for the
$L=5$ case. An example of this neural network can be found in the
notebook ~\texttt{AutoEncoder.ipynb} in Ref.~\cite{git-ref}.

\begin{center}
\begin{table}
\centering
\begin{tabular}{cccccc}
\hline
LayerName & Input & LayerType & Activation & Nb of units & Kernel \\
\hline
\textbf{\underline{AutoEncoder}} \vspace{0.2cm} \\ 
CSq1 & J & conv & ReLu & $32$ & $ 3 \times 3$ \\
CL1 & J & conv & ReLu & $16$ & $ L_x \times 1$  \\
CC1 & J & conv & ReLu & $16$ & $ 1 \times L_y$ \\
LatentRepr & [CSq1,CL1,CC1] & FF & ReLu & $50$ ($= 5.5.2$) & \\
ConvDec1 & LatentRepr & conv & ReLu & $64$ & $3 \times 3$ \\
UpS & ConvDec1 & UpS & & & $2 \times 2$ \\
CSq2 & UpS & conv & ReLu & $32$ & $ 3 \times 3$ \\
CH2 & UpS & conv & ReLu & $32$ & $ L_x \times 1$ \\
CV2 & UpS & conv & ReLu & $32$ & $ 1 \times L_y$ \\
ConvDec & [CSq2,CH2,CV2] & conv & Linear & $1$ & $5 \times 5$ \\
\hline
\end{tabular}
\caption{A typical architecture used for the autoencoder, FF stands for feed-forward and UpS for upsampling, conv for convolutional, ReLu for Rectified Linear unit.}
\label{tab:archAE}
\end{table}
\end{center}

\subsection{Learning} 

The learning procedure was performed by using a linear activation for
the last layer, together with a Minimum Square Error (MSE) loss
function on all the nodes of the system. The MSE is computed between
output of the autoencoder for the input $\boldsymbol{J}$, and its
comb-gauge representative $\mathrm{G}(\boldsymbol{J})$. In principle, it would
be possible to use as loss function a binary cross entropy, together
with a $\tanh$ for the activation function, taking advange of the binary nature of the couplings. However, we did not find
any improvement when using these parameters w.r.t. the others.  We note as well that,
because we use the chess transformation, the loss is defined on all
the pixels, including the dummy ones. Neglecting dummy pixels,
however, did not result in any improvement.

\subsection{Tests} 

It is known that DNNs are prone to overfit the dataset. Hence, in
order to be sure that the autoencoder did learn a general property, we perform several checks on a test set (i.e. a set of orbits not used to train the network) on our well trained machine. In general, we compare the output of the autoencoder (the reconstructed comb gauges) for two distinct input samples $\{\V{J},\V{J^\prime}\}$. The comparison is done by counting the number of different couplings. We consider four diverse situations:
\begin{enumerate}
    \item The two samples are from the same gauge orbit, i.e. $\V{J^\prime}=\mathrm{O}(\V{J})$.
    \item Two samples separated by a line and a gauge transformation, i.e. $\V{J^\prime}=\mathrm{O}(L(\V{J}))$.
    \item Two samples separated by a random-link and a gauge transformation, i.e. $\V{J^\prime}=\mathrm{O}(\mathrm{R}_q(\boldsymbol{J}))$.
    \item Two random samples.
\end{enumerate}
\noindent We show in Table \ref{tab:resL5} the results of these comparisons averaged over 1000 pairs of each situation. Outputs from samples in the same orbit are essentially equal (only a $\sim 3\%$ of the couplings are different). If the gauge-fixing were perfect, they should be strictly equal. However, a much larger difference is observed in the outputs of the rest of the cases. Notwithstanding, we would like to stress that we needed a large number of samples to be able to distinguish case no.1 from no.2. With a fewer numbers, outputs of test no.2  were essentially equal.

\begin{table}[!h]
    \centering
    \begin{tabular}{c|c|c|c}
         Same Orbit & Diff. Orbit (Line) & Diff. Orbit ($q=0.1$)  & Random \\
         \hline
         $\sim 3$ \% &   $\sim 21 \%$ & $\sim 30 \%$ & $\sim 50\%$
    \end{tabular}
    \caption{Results for the autoencoder for the size $L=5$. We observe a clear gap  when samples came from the same orbit gauge with respect to even a small alteration (such as flipping a small fraction of coupling or a line).}
    \label{tab:resL5}
\end{table}
\begin{figure}[!b]
    \centering
    \includegraphics[scale=0.44,trim=50 30 40 0 ,clip]{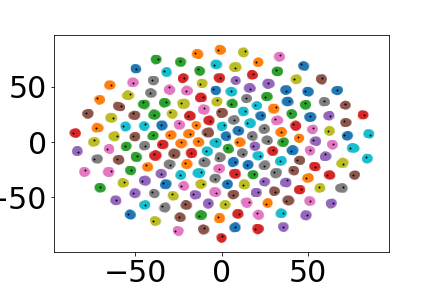}
    \includegraphics[scale=0.44,trim=50 30 40 0 ,clip]{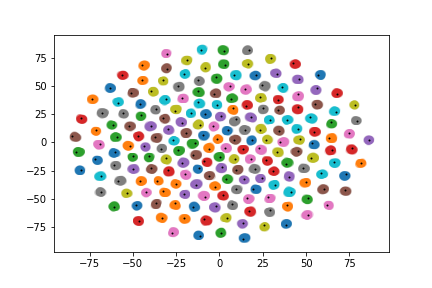}
    \includegraphics[scale=0.44,trim=50 30 40 0 ,clip]{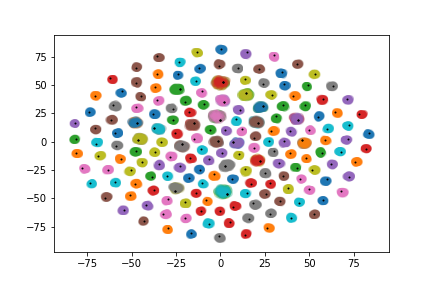}
    \caption{On the left, the t-sne representation for $N_\mathrm{s}=20000$, each one being a gauge transform from $200$ randomly chosen orbits. On the middle, $100$ orbits are chosen randomly whereas $100$ others orbits are constructed by operating a $\mathrm{R}_q$ transformation (with $q=0.1$) from the first hundred ones. On the right, the same but applying a line transformation $\mathrm{L}(\V{J})$. The last two cases are much more difficult. In the first case, the clusters are all well-separated, yet, in the last case, a very few mistakes still occur.}
    \label{fig:lat-rep}
\end{figure}

We add an additional test on the trained the network. We want to understand if the network manages to learn an (almost) unique representation for a given orbit. To do that, we use the $t-sne$ representation to project in two dimensions the  high-dimensional latent space. If the network is able to cluster well the samples in distinct orbits (that is, if the network learned the gauge symmetry), the $t-sne$ transformation of different orbits should be well-separated. On Fig. \ref{fig:lat-rep} we illustrate the clustering generated by our trained autoencoder (for $L=5$) using as input the following group of test sets of $N_\mathrm{s}=200000$ samples each: 
\begin{enumerate}
    \item We generate $N_{\mathrm{g}}=200$ random orbits $\V{J}$, and take $100$ gauge transformations from each $\boldsymbol{J'} = \mathrm{O}(\boldsymbol{J})$.
    \item We generate $N_{\mathrm{g}}=100$ random orbits $\V{J}$, and another $100$ orbits constructed applying the $\mathrm{R}_{q=0.1}(\V{J})$ transformation to the $100$ random ones. Again, we take $100$ gauge transformations from each orbit.
    \item We generate $N_{\mathrm{g}}=100$ random orbits $\V{J}$, and another $100$ orbits constructed by applying the $\mathrm{L}(\V{J})$ transformation on the $100$ random ones. Again, we take $100$ gauge transformations from each orbit.
\end{enumerate}
\noindent We show the result of the $t-sne$ two-dimensional representations of these three groups on Fig. \ref{fig:lat-rep}. We clearly see very good clustering properties for all the groups, though the third case remains sometimes difficult.

\subsection{Classifier based on the latent representations}

Not very surprisingly, one can also train a  neural-network to tell us whether two latent representations (generated by our trained autoencoder using two different samples) belong to the same gauge orbit or not, thus doing the job of our previous classifier (discussed in Section~\ref{sec:class}). To do so, we concatenate the two latent representations and feed them to various CNNs and a classification layer. Various architectures worked there, we put one as an example in the notebook ~\texttt{AutoEncoder.ipynb} in Ref.~\cite{git-ref}, whose details are reproduced on  Table~\ref{tab:ae-class}. When the autoencoder is well-trained, the classifier quickly reaches an accuracy above $98\%$.

\begin{table}[!h]
\centering
\begin{tabular}{cccccc}
\hline
LayerName & Input & LayerType & Activation & Nb of units & Kernel \\
\hline
\textbf{\underline{Enc-Classif}} \vspace{0.2cm} \\ 
Concat & [LatentRepr($J_1$),LatentRepr($J_2$)]& & & & \\
Conv1 & Concat & conv & ReLu & $16$ & $32$ \\
MaxP1 & Conv1 & pooling & MaxPooling & & $2$ \\
Conv2 & MaxP1 & conv & ReLu & $32$ & $16$ \\
MaxP1 & Conv2 & pooling & MaxPooling & & $2$ \\
D1 & MaxP1 & FF & ReLu & 32 & \\
Out & D1 & FF &Softmax & 2 & \\
\hline
\end{tabular}
\caption{Architecture used for the classifier of latent representations (created by the autoencoder). FF stands for feed-forward and UpS for upsampling.}
\label{tab:ae-class}
\end{table}

\twocolumngrid
\bibliography{biblioSL}

\end{document}